\documentclass[aps,prb,superscriptaddress,floatfix,twocolumn]{revtex4-2}
\usepackage{graphicx}
\usepackage{dcolumn}
\usepackage{bm}

\usepackage{float}
\usepackage{siunitx}
\usepackage{subfigure}
\usepackage{xcolor}
\usepackage[utf8]{inputenc}
\usepackage[T1]{fontenc}
\usepackage{siunitx}
\usepackage{mathptmx}
\usepackage{etoolbox}
\usepackage{amsmath}

\makeatletter
\def\@email#1#2{%
 \endgroup
 \patchcmd{\titleblock@produce}
  {\frontmatter@RRAPformat}
  {\frontmatter@RRAPformat{\produce@RRAP{*#1\href{mailto:#2}{#2}}}\frontmatter@RRAPformat}
  {}{}
}%
\makeatother

\begin{document}

\preprint{AIP/123-QED}

\title{Modification of the scattering mechanisms in bilayer graphene in proximity to a molecular thin film probed in the mesoscopic regime}

\author{Anise Mansour} 
\affiliation{Department of Physics and Astronomy, California State University Long Beach, Long Beach, California 90840, USA}
\altaffiliation{AM, DD and MKDDM contributed equally to this work.}

\author{Deanna Diaz} 
\affiliation{Department of Physics and Astronomy, California State University Long Beach, Long Beach, California 90840, USA}
\altaffiliation{AM, DD and MKDDM contributed equally to this work.}

\author{Movindu K. D. Dissanayake Mudiyanselage}
\affiliation{Department of Physics and Astronomy, California State University Long Beach, Long Beach, California 90840, USA}
\altaffiliation{AM, DD and MKDDM contributed equally to this work.}

\author{Erin Henkhaus} 
\affiliation{Department of Physics and Astronomy, California State University Long Beach, Long Beach, California 90840, USA}

\author{Jungyoun Cho} 
\affiliation{California NanoSystems Institute (CNSI), University of California, Los Angeles, California 90095, USA}

\author{Vinh Tran} 
\affiliation{Department of Physics and Astronomy, California State University Long Beach, Long Beach, California 90840, USA}


\author{Francisco Ramirez} 
\affiliation{Department of Physics and Astronomy, California State University Long Beach, Long Beach, California 90840, USA}

\author{Eric Corona-Oceguera} 
\affiliation{Department of Physics and Astronomy, California State University Long Beach, Long Beach, California 90840, USA}

\author{Joshua Luna} 
\affiliation{Department of Physics and Astronomy, California State University Long Beach, Long Beach, California 90840, USA}

\author{Kenta Kodama} 
\affiliation{Department of Physics and Astronomy, California State University Long Beach, Long Beach, California 90840, USA}

\author{Yueyun Chen} 
\affiliation{Department of Physics and Astronomy, University of California, Los Angeles, California 90095, USA}
\affiliation{California NanoSystems Institute (CNSI), University of California, Los Angeles, California 90095, USA}

\author{Ho Chan} 
\affiliation{Department of Physics and Astronomy, University of California, Los Angeles, California 90095, USA}
\affiliation{California NanoSystems Institute (CNSI), University of California, Los Angeles, California 90095, USA}

\author{Jacob Weber} 
\affiliation{Department of Physics and Astronomy, California State University Long Beach, Long Beach, California 90840, USA}

\author{Blake Koford} 
\affiliation{Department of Physics and Astronomy, California State University Long Beach, Long Beach, California 90840, USA}

\author{Patrick Barfield} 
\affiliation{Department of Physics and Astronomy, California State University Long Beach, Long Beach, California 90840, USA}

\author{Maya Martinez} 
\affiliation{Department of Physics and Astronomy, California State University Long Beach, Long Beach, California 90840, USA}

\author{Kenji Watanabe} 
\affiliation{Research Center for Electronic and Optical Materials, National Institute for Materials Science, 1-1 Namiki, Tsukuba 305-0044, Japan}

\author{Takashi Taniguchi} 
\affiliation{Research Center for Electronic and Optical Materials, National Institute for Materials Science, 1-1 Namiki, Tsukuba 305-0044, Japan}

\author{B. C. Regan} 
\affiliation{Department of Physics and Astronomy, University of California, Los Angeles, California 90095, USA}
\affiliation{California NanoSystems Institute (CNSI), University of California, Los Angeles, California 90095, USA}

\author{Matthew Mecklenburg} 
\affiliation{California NanoSystems Institute (CNSI), University of California, Los Angeles, California 90095, USA}

\author{Thomas Gredig} 
\affiliation{Department of Physics and Astronomy, California State University Long Beach, Long Beach, California 90840, USA}

\author{Claudia Ojeda-Aristizabal}
\affiliation{Department of Physics and Astronomy, California State University Long Beach, Long Beach, California 90840, USA}
\email[Corresponding author~]{Claudia.Ojeda-Aristizabal@csulb.edu}

\date{\today}

\begin{abstract}
   Quantum coherent effects can be probed in multilayer graphene through electronic transport measurements at low temperatures. In particular, bilayer graphene is known to be susceptible to quantum interference corrections of the conductivity, presenting weak localization at all electronic densities, and dependent on different scattering mechanisms as well as on the trigonal warping of the electron dispersion near the K and K' valleys. Proximity effects with a molecular thin film influence these scattering mechanisms, which can be quantified through the known theory of magnetoconductance for bilayer graphene. Here, we present weak localization measurements in a copper-phthalocyanine / bilayer graphene / h-BN heterostructure that suggest an important suppression of trigonal warping effects in bilayer graphene (BLG), restoring the manifestation of the chirality of the charge carriers in the localization properties of BLG. Additionally, we observe a charge transfer of 3.6$\times$10$^{12}$cm$^{-2}$ from the BLG to the molecules, as well as a very small degradation of the mobility of the BLG/h-BN heterostructure upon the deposition of copper phthalocyanine (CuPc). The molecular arrangement of the CuPc thin film is characterized in a control sample through transmission electron microscopy, that we relate to the electronic transport results.   
\end{abstract}


\maketitle

\section{Introduction}
In the mesoscopic regime, that is, when the dimensions of a system are small enough to be sensitive to quantum effects and large enough to take into account the contribution of multiple electrons, the quantum interference of electronic trajectories leads to a correction of the conductance that is intimately linked to the character of the charge carriers. In graphene, carriers are chiral, their pseudo-spin, related to the existence of two atoms per unit cell, is either parallel or anti-parallel to the momentum. This makes that if an electron in graphene completes a closed trajectory, the electron wavefunction will acquire an additional phase of $\pi$, leading to a destructive interference of the electronic wavefunctions when the electrons propagate in opposite directions around a trajectory \cite{PhysRevLett.103.226801}. In contrast, the carriers' wave function in bilayer graphene (BLG) while also being chiral, have a geometric phase of 2$\pi$ leading instead to a constructive interference for similar interfering electronic trajectories. This phenomenon in a BLG sample increases the probability of electrons scattering back and therefore decreases the probability of reaching the drain electrode of the sample, resulting in a decrease of electrical conductance with respect to its classical value in the frame of the Drude model \cite{PhysRevLett.98.176805}. This effect is known as weak localization and can be detected by applying a small magnetic field perpendicular to the sample, that adds an additional phase to the interfering electronic wavefunctions and destroys the interference, resulting in a positive magnetoconductance. In BLG, weak localization is different from the one present in conventional 2-dimensional systems in that it is sensitive to not only inelastic processes that break the phase coherence but also, to elastic scattering events such as intervalley scattering, associated to sharp defects that are able to scatter electrons between the two valleys of BLG. Furthermore, weak localization in BLG is affected by the trigonal warping of the energy spectrum of the carriers \cite{PhysRevLett.98.176806, PhysRevLett.98.176805}.

The presence of a molecular thin film is able to influence these quantum interference effects. Previous works on monolayer graphene covered with Pt-porphyrines have found that the magnetic moment of these molecules, tunable with the vertical electric field imposed by a gate voltage, has a profound impact on quantum coherent phenomena in graphene, such as universal conductance fluctuations and superconducting proximity effects \cite{PhysRevB.93.045403}. Time reversal symmetry breaking brought by the magnetic moment in these molecules creates an odd component in the conductance fluctuations as well as a suppression of the supercurrent in superconductor/graphene/superconductor junctions. Other works on graphene decorated with adatoms have also found signature of modified quantum interference effects, that points to a decrease in weak localization due to the presence of the adatoms and an increase of intervalley scattering, despite the expected stronger long-range intravalley Coulomb scattering from the adatoms. It has also been shown that the presence of the adatoms significantly deteriorates the electronic mobility of the original device \cite{PhysRevB.91.245402, PhysRevB.95.075405}.  

    \begin{figure}
    \includegraphics[width=0.45\textwidth]{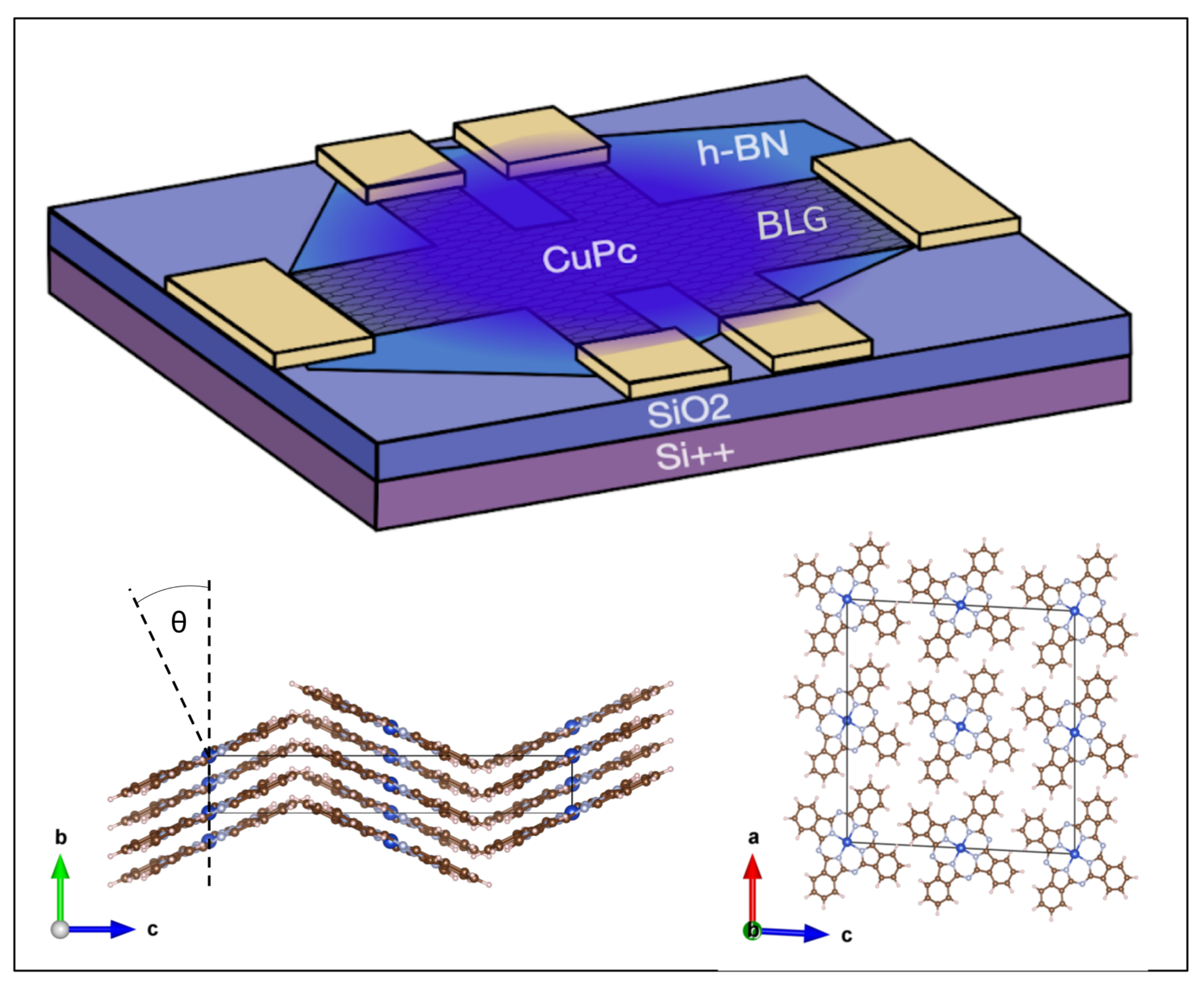}\caption{Top: Schematics of the CuPc/BLG/h-BN device. Bottom: Crystal Structure of CuPc. The angle $\theta$ defines the different existing polymorphs.}
    \label{fig:CuPc}
    \end{figure}

Here, we study the electronic transport of a copper-phthalocyanine (CuPc)/BLG/hexagonal boron nitride (h-BN) heterostructure in the mesoscopic regime. Phthalocyanines are macrocyclic planar aromatic molecules that admit at their center any transition metal from the periodic table, making them highly versatile molecules (see Figure \ref{fig:CuPc}). CuPc is the most studied phthalocyanine, with a copper ion that has an unpaired electron spin, making it a paramagnetic metal-organic molecule of spin angular momentum s=1/2 \cite{Bartolomé2014}. In the bulk, the copper atoms of CuPc have tendency to form one-dimensional chains along the b-axis, as a result of van der Waals molecule-molecule interactions being stronger when the molecules are face-to-face rather than side-to-side. There are different polymorphs of CuPc, distinguished by the angle $\theta$ between the b-axis and the normal to the plane of the molecule. In the form of a thin film, CuPc grows in a lying configuration for most metallic substrates. When deposited at room temperature, CuPc tends to form a metastable phase known as the $\alpha$ phase, where $\theta\approx 25\textdegree$ (see Figure \ref{fig:CuPc}) \cite{Bartolomé2014}.      

Our measurements suggest that the presence of CuPc has the effect of increasing the weak localization on the BLG/h-BN heterostructure. A fit to the theory for BLG allows us to identify longer trigonal warping characteristic scattering times, that we associate to a reduction of the trigonal warping in BLG due to the presence of the CuPc molecules. Additionally, we observe an important charge transfer from the BLG to the CuPc molecules and a reasonable preservation of the mobility of the BLG/h-BN heterostructure upon the deposition of the molecules.  



\section{Sample Fabrication} 
\begin{figure}
    \includegraphics[width=0.5\textwidth]{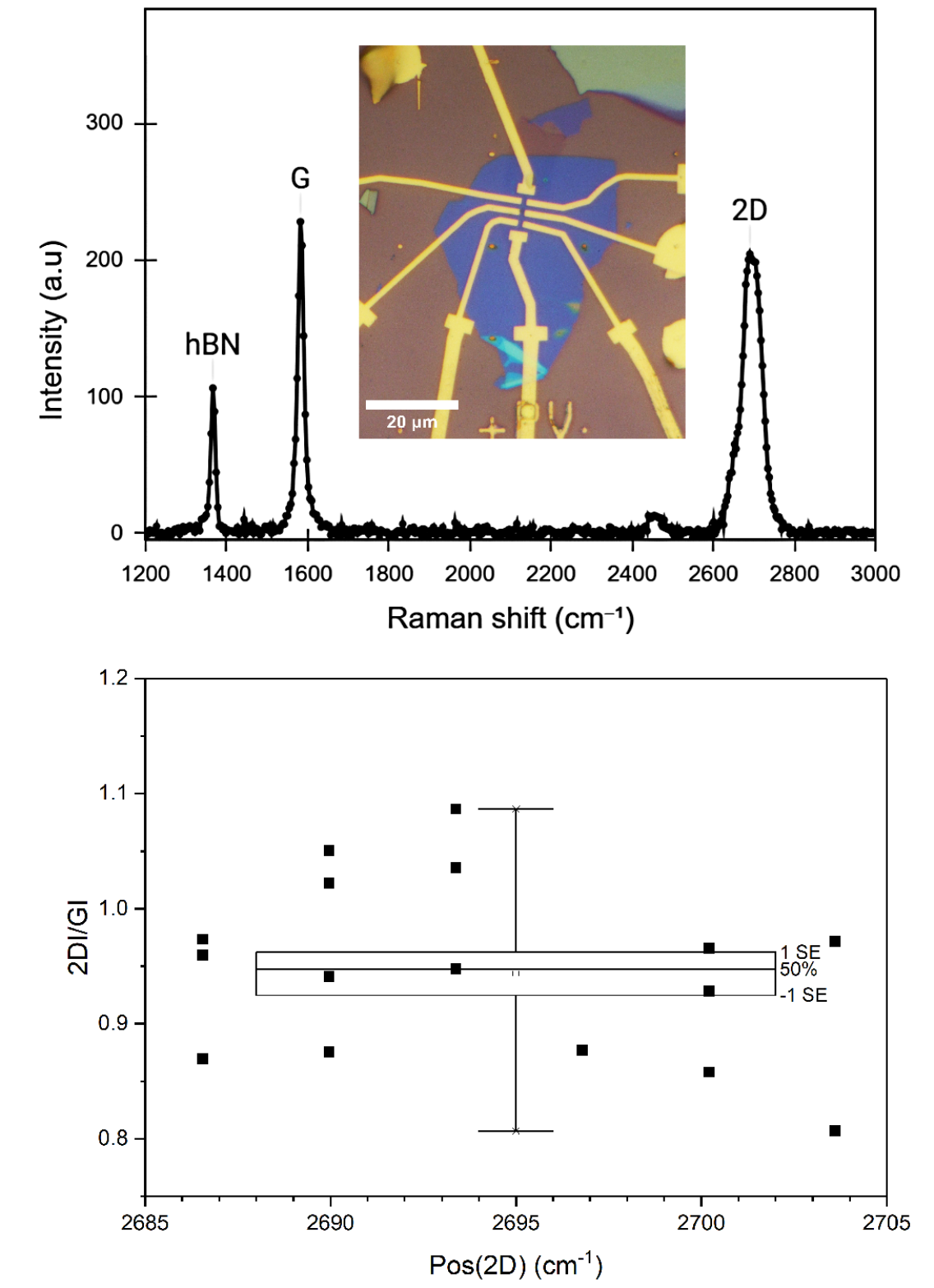}\caption{Top: Raman spectra of graphene on hBN. The inset shows an optical image of the BLG/h-BN device. Bottom: Ratio of the 2D peak intensity to the G peak intensity plotted against the 2D peak Raman shift, with the mean (white square), median (50\%), and standard error of the mean ($\pm$1 SE) indicated. Black squares represent individual measurements at different locations of the sample.}
    \label{fig:Gr/hBN Raman}
    \end{figure}

    A heterostructure of BLG/h-BN was fabricated using the Zomer method \cite{10.1063/5.0146141}. BLG crystals were exfoliated on elvacite coated glass slides using Scotch tape, while hexagonal boron nitride (h-BN) crystals were exfoliated on Si/SiO\textsubscript{2} wafers. After an acetone, methanol, and IPA wash, visible glue residue remained on the hBN, necessitating further annealing in a CVD furnace. The Si/SiO\textsubscript{2} wafers with exfoliated h-BN were annealed in a hydrogen(10 scc/m) / argon(190 scc/m) flow while heating up the sample to 350°C over 2 hours in a quartz tube. The temperature was maintained at 350°C for 3 hours to ensure the removal of residues, before cooling down the furnace to room temperature over a period of 3.25 hours.
    The BLG and hBN crystals were then aligned and transferred on top of each other at $ \approx$ 80 \textdegree C until the elvacite melted on the wafer. The wafer and slide were cooled, and the elvacite was dissolved in dichloromethane, leaving behind the BLG/hBN heterostructure on the Si/SiO\textsubscript{2} wafer. The sample was characterized through Raman spectroscopy using a Witec Alpha 300R spectrometer with a 532 nm laser. It has been reported that the relative intensity of the 2D to G peak ratio significantly varies with graphene layer number, ranging from 2.3 to 2.8 for single-layer graphene (SLG), 0.86 to 0.94 for bilayer graphene (BLG), and 0.63 to 0.74 for trilayer graphene (TLG) \cite{FATES2019390}. We measured a ratio of the 2D to G peak intensity within a range of 0.8–1.1, as shown in Figure \ref{fig:Gr/hBN Raman} demonstrating that our sample is a BLG.
    
    \color{black}The BLG was coated with PMMA 495 and PMMA 950 in order to define electrode patterns through electron beam lithography. Titanium and gold (10 nm/40 nm) were later deposited via electron beam evaporation.

\begin{figure*}
\includegraphics[width=\textwidth]{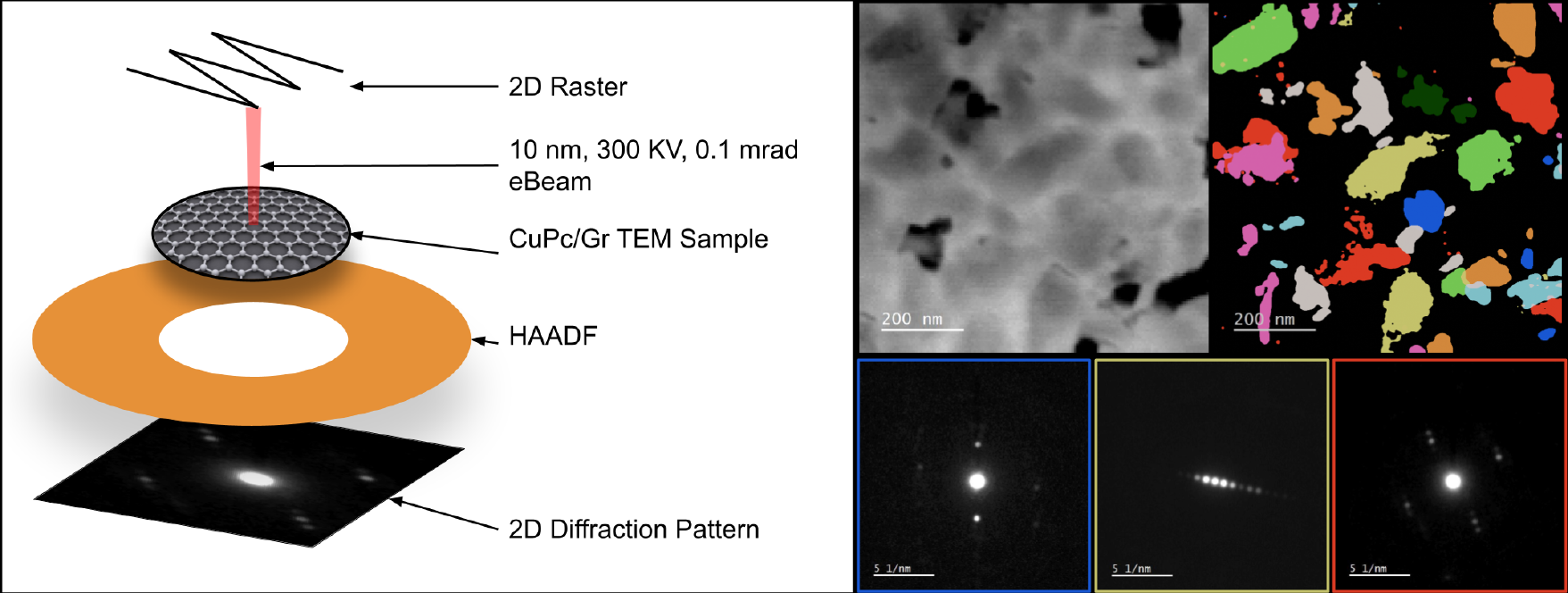}\caption{Left: 4D-STEM Schematic showing the electron beam being rastered across the CuPc-coated graphene TEM grid, the high-angle annular dark field (HAADF) detector and the camera detector at the bottom, that collects the diffraction patterns. Right: Virtual bright-field image. The color-coded grain map of the measured CuPc/graphene sample is shown. The diffraction patterns of some of the grains are shown, framed in a color corresponding to the one used for certain grains.}
\label{fig:4DSTEM}
\end{figure*}
    
Transport measurements were conducted on the device prior to the deposition of CuPc, as will be detailed later. The CuPc was thermally evaporated using a commercially available high purity powder of CuPc (Ted Pella) using an organic materials- dedicated thermal evaporator. The deposition was performed at 0.3\AA/s until reaching a thickness of 22 nm in a base pressure of 1$\times 10^{-6}$ Torr. The sample was kept at room temperature during the deposition process. Simultaneously, CuPc was deposited on a control sample, made of suspended graphene on lacey carbon, compatible with transmission electron microscopy experiments.

\section{Sample characterization through 4D Scanning transmission electron microscopy}
Transmission Electron Microscopy (TEM) is an imaging technique that utilizes high-energy electrons to probe thin samples, generating images based on the electron-sample interactions. Unlike conventional TEM, which uses a broad electron beam, Scanning Transmission Electron Microscopy (STEM) employs a finely focused electron probe. This reduction in spot size results in increased electron fluence, allowing for high-resolution imaging and diffraction data acquisition at specific sample locations. 4D-STEM measures the 2D diffraction pattern at each position of a 2D STEM map, creating a 4D data set.
We chose this diffraction space imaging over real-space imaging as it enables higher resolution for a given electron dose, minimizing radiation damage to beam-sensitive samples such as phthalocyanine molecules. 4D-STEM allowed us to resolve individual CuPc crystals in our control sample of CuPc/graphene.

Figure \ref{fig:4DSTEM} illustrates the 4D-STEM measurement. A small electron beam is rastered across a CuPc-coated graphene TEM grid. Live 2D spatial images are first captured using the high-angle annular dark field (HAADF) detector, followed by sequential collection of diffraction patterns using a camera detector at the bottom. By using virtual selected and objective apertures to isolate signals across multiple probe positions and diffraction peaks, individual Bragg peaks in reciprocal space were identified and correlated with real-space imaging to construct a composite image of the grain morphology (see Figure \ref{fig:4DSTEM}). The analysis of this image allowed us to calculate an average grain size of the CuPc thin film in our control sample of $42\pm0.7$ nm. This is comparable to the grain size reported for FePc deposited at room temperature, which can be tuned by adjusting the temperature of the substrate during deposition \cite{PhysRevB.80.174118}.

The TEM control sample was fabricated simultaneously with the BLG/h-BN device, sharing the same deposition conditions. We can therefore assume a similar CuPc grain distribution in the CuPc/BLG/h-BN transport device. Understanding the molecular arrangement of the molecules on BLG/h-BN is important, as the size of the crystalline domains can potentially influence the charge transfer and addition of scatterers to the original device, as will be developed later.

\section{Electronic transport measurements at low temperatures}
    
\begin{figure}
\includegraphics[width=0.45\textwidth]{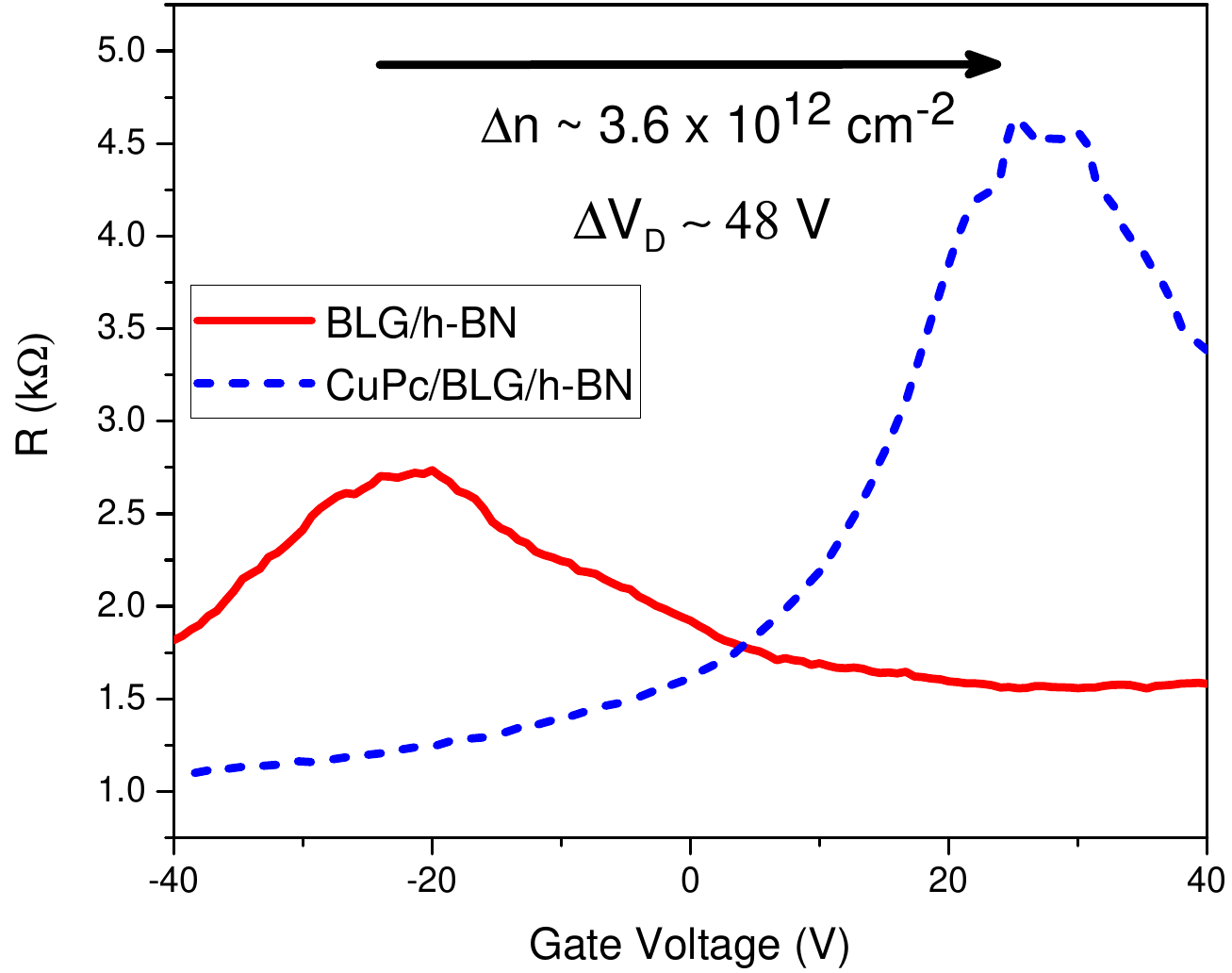}
\caption{Backgate dependence of the resistance of the initial BLG/h-BN device (red) and after the deposition of CuPc (blue). Upon the deposition of the molecules, the charge neutrality point initially at -24 V sifts to 24 V, indicating a charge transfer of \SI{3.6e12}{\per\centi\meter\squared}}. The data was taken on a square junction of length 4.5$\times 10^{-6}$ m.
\label{fig:charge_transfer}
\end{figure}

We characterized our BLG/h-BN transport device through electronic transport measurements at low temperatures before the deposition of the CuPc molecules. Experiments were performed in a closed cycle cryostat with a superconducting magnet, where the differential resistance dV/dI in a two-probe setup was measured by modulating a bias current of typical amplitude 10-100 nA and measuring the voltage drop through standard lock-in detection. A measurement of the backgate voltage dependence of the resistance showed that our device was initially electron doped. Upon deposition of the CuPc molecules, we observed an important shift of the Dirac point, of about 48 V towards the positive voltages, indicating the hole doping of the BLG/h-BN device and a transfer of electrons from the BLG/h-BN to CuPc of $(3.6\pm 0.05)\times 10^{12}$ cm$^{-2}$ (see Figure \ref{fig:charge_transfer}). The uncertainty on the charge transfer was calculated from the error of the fits presented in the Supplementary Materials, that allow to determine the gate voltage value of the charge neutrality point V$_D$. The uncertainty for V$_D$, $\Delta V_D=\pm 0.7$ V affects all quantities that depend on the gate voltage or the electronic density, presented in this section. The value found for $\Delta V_D$ coincides with the gate voltage step used for the back gate-dependent resistance measurements. 

The previously mentioned electron transfer from graphene to CuPc is equivalent to the transfer of $67\pm1$ holes per CuPc grain to the BLG/h-BN heterostructure. This is translated in a Fermi energy shift of $120\pm2$ meV in the BLG, calculated using the relation between the Fermi energy and the Fermi wave vector, that for bilayer graphene is $E=\frac{\hbar^2k_F^2}{2m^*}$, with $k_F=\sqrt{n\pi}$ . 
This important charge transfer can be understood through the work function values of the materials involved. Graphene's work function on SiO$_2$ is known to be 4.6 eV, being slightly larger for BLG ($\approx 4.7$ eV) \cite{doi:10.1021/nl901572a}. Recent works that use Kelvin probe force microscopy have found a work function of 4.6 eV for quasi-free standing BLG \cite{Melios2015-ld}, that represents a closer situation to a BLG/h-BN heterostructure, where the BLG is isolated from the SiO$_2$ charge impurities. For CuPc in the form of a thin film, scanning electron microscopy experiments report a work function of $\approx 4.9$ eV \cite{VIJAYAN2018348}. These values explain CuPc acting as an electron acceptor in our CuPc/BLG/h-BN device. Similarly to CuPc, FePc has been reported to be an electron acceptor when in contact to graphene \cite{1627892824705-1270331098}, while vanadyl-Pc on graphene acts as an electron donor \cite{doi:10.1021/jp200386z}.

Surprisingly, the deposition of the CuPc thin film on the BLG/h-BN device creates only a small degradation of the mobility and mean free path of the original device, demonstrating that the CuPc thin film has a minimal effect on the landscape of long and short range scatterers that limit the mobility in graphene. This is in contrast to adatoms, that while able to offer new functionalities to graphene, are highly detrimental to its mobility \cite{PhysRevB.83.085410, PhysRevB.82.073403, PhysRevB.86.075433, PhysRevB.95.075405, PhysRevB.91.245402}.
We use the Drude formula to calculate the mobility of our device $\mu=\frac{\sigma}{en}$ and the capacitor model to estimate the electronic density at different gate voltages, $n=\frac{\epsilon_0\epsilon_r}{ed}(V_g-V_D)$, with V$_D$ the gate voltage of the Dirac point (see Supplementary Materials). We deduce an electron mobility for our BLG/h-BN device of $3200\pm 200$ cm$^2$/Vs away from the Dirac point (at 1$\times$10$^{12}$ cm$^{-2}$ carrier density) before the deposition of CuPc, becoming $2800\pm 100$ cm$^2$/Vs after the deposition of the molecules. We also calculated the mean free path at the electron density 1$\times$10$^{12}$ cm$^{-2}$ from the diffusion coefficient $D\equiv v_F l_e/2$, where $v_F=\hbar k_F/m^*$ for bilayer graphene. The diffusion coefficient is related to the conductivity through Einstein's relation $D=\frac{\sigma}{e^2\rho(E_F)}$ \cite{Datta1995-it}. Using the density of states at the Fermi level for bilayer graphene $\rho(E_F)=\frac{2m^*}{\pi\hbar^2}$ yields $l_e=\frac{h\sigma}{2k_Fe^2}$, an expression that is in fact independent on the dispersion relation of the two-dimensional system \cite{Akkermans2004-ir}. We find an electron mean free path for electrons away from the Dirac point of $37\pm1$  nm, that becomes $32\pm1$ nm after the deposition of the molecules (see Figure \ref{fig:Mu_mfp} in Supplementary Materials). The calculation of the mean free path allows as to estimate k$_F$l$_e$, that measures the effect of disorder on the electronic wave functions. k$_F$l$_e$ >> 1 corresponds to a weakly disordered system and a good conductor, while k$_F$l$_e$ << 1 is characteristic of an insulator with strongly localized electronic wave functions \cite{Akkermans2004-ir}. This value changes from $6.7\pm0.3$ for the BLG/h-BN device to $5.7\pm0.2$ after the deposition of CuPc, measured at 1$\times$10$^{12}$ cm$^{-2}$ electron carrier density. The initial device was therefore at an intermediate regime, as expected for graphene, where the disorder plays an important role on quantum interference effects, affecting the measured conductivity. The small change in the value of k$_F$l$_e$ shows that the addition of the molecules lightly increases the disorder in the device, leading to somewhat more localized wave functions. As previously mentioned, the conductivity can be expressed in terms of k$_F$l$_e$ ($\sigma=(2e^2/h)l_ek_F$). The change in k$_F$l$_e$ can also be deduced when inspecting the conductivity in units of 2e$^2$/h (see figure \ref{fig:sigma} in Supplementary Materials). 

By plotting the conductivity, we can also quantify the increase of the width of the Dirac point, from 8.6$\pm$0.4 V in the BLG/h-BN device to 10$\pm$0.4 V after the deposition of the CuPc molecules (see Figure \ref{fig:sigma} from the Supplementary materials). This corresponds to a change from $(0.65\pm0.05)\times 10^{12}$/cm$^2$ to $(0.76\pm0.05)\times 10^{12}$/cm$^2$. Such result can be understood within the approach of Adam, Das Sarma et al. \cite{Adam2007-fn} as an increase of the residual electronic density. In their approach, the main scattering mechanism at low carrier densities comes from charged long-range impurities that create a non-uniform electron-hole puddle landscape \cite{Zhang2009-ip}, understood as the main cause of a minimum conductivity plateau in graphene \cite{Chen2008-dt}. Therefore within this frame, CuPc molecules add some few charge impurities to the original BLG/h-BN device, $(0.11\pm0.05)\times10^{12}$/cm$^2$, corresponding to $2\pm1$ charged impurities per CuPc grain.

We can also estimate the elastic scattering time $\tau_e=l_e/v_F$, the momentum relaxation time, given by an individual collision destroying the initial momentum of the electrons \cite{Datta1995-it}. We find $\tau_e=(6.0\pm0.2)\times 10^{-14}$ s, coincidentally, very close to the value measured in the past at the same electron density (1$\times10^{12}$/cm$^2$) for a BLG through magnetoresistance measurements \cite{PhysRevLett.104.126801, Monteverde2012-io}. The deposition of the molecules change this time to $\tau_e=(5.2\pm0.2)\times 10^{-14}$ s.      

\begin{figure*}
\includegraphics[width=\textwidth]{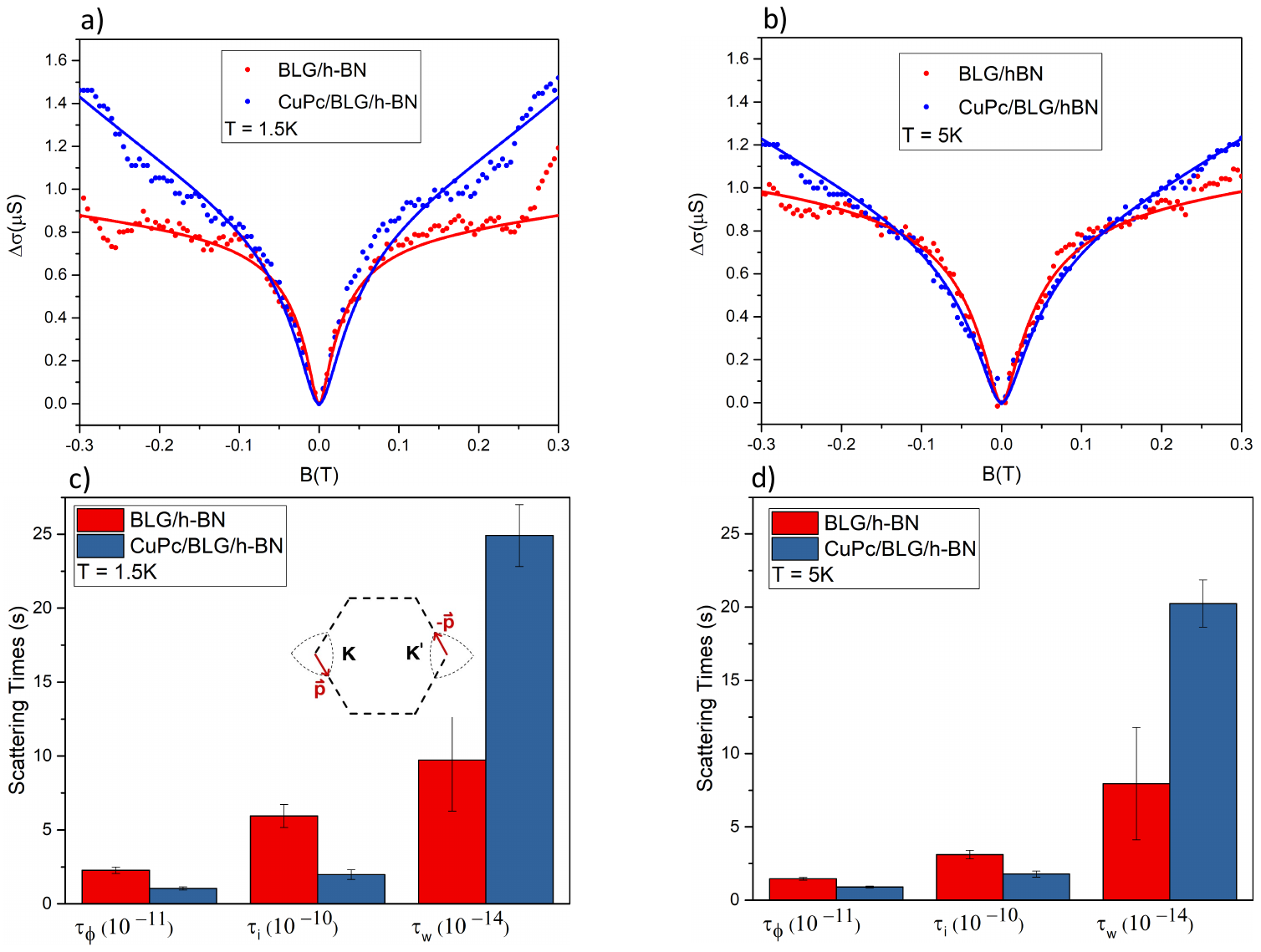}
\caption{Magnetic field dependence of the conductivity measured on the original BLG/h-BN device (red) and after the CuPc deposition (blue) at 1.5K (a) and 5K (b), at electronic densities of (0.7$\pm 0.05)\times 10^{12}$ cm$^{-2}$. The continuous lines represent the fitting to equation \ref{WL_BLG}, performed using a nonlinear least squares algorithm, \texttt{scipy.optimize.curve\_fit} from Python. The fit was performed over the negative fields and mirrored to the positive fields. (c) and (d) represent the obtained characteristic times for each temperature. The error bars correspond to the standard error for each fitted parameter. The data was taken on a rectangular junction of length 0.98 $\mu$ m and width 5.7 $\mu$ m. The inset represents BLG's first Brillouin zone and the Fermi lines close to two inequivalent valleys \textbf{K} and \textbf{K'}. Trigonal warping creates a \textbf{p}\textrightarrow-\textbf{p} asymmetry of the dispersion at each valley (see text).} 
\label{fig:WL}
\end{figure*} 

The preserved mobility and mean free path after the deposition of CuPc is consistent with previous reports on other molecules in graphene, such as C$_{60}$ \cite{doi:10.1021/acsnano.7b00551} and Pt-porphyrines \cite{PhysRevB.93.045403}. The latter are similar to the phthalocyanines, with a macrocyclic structure with delocalized $\pi$ electrons, similar to CuPc. Pt-porphyrines have been reported not only to preserve, but in some cases to improve the mobility of graphene upon the deposition of the molecules \cite{PhysRevB.93.045403}. Similar results have been found with other planar molecules on graphene \cite{Datta2017-zl}. It is worth pointing out that in the works with planar molecules, the molecules are dispersed in a solution and drop casted on the graphene device, and in some cases annealed to remove the solvent \cite{Datta2017-zl}, a process that can potentially modify the mobility of the original device. Additionally, the crystallinity of the deposited molecules is unknown. This is in contrast to the CuPc molecules in the present work, that are deposited through a low temperature thermal evaporation while the BLG/h-BN remains at room temperature, and where a control TEM-compatible device gives information about the arrangement of the molecules on the device.  

         
We now turn to the analysis of the weak localization in the BLG/h-BN before and after the deposition of the CuPc thin film. In bilayer graphene, the quantum correction to the Drude conductivity, as a function of magnetic field is given by \cite{PhysRevLett.98.176805}: 
   
    \begin{align}
    \Delta \sigma (B) &= \frac{e^2}{\pi h} \left[ F\left( \frac{B}{B_\phi} \right) 
    - F\left( \frac{B}{B_\phi + 2B_i} \right) 
    + 2 F\left( \frac{B}{B_\phi + B_*} \right) \right], \nonumber \\
    F(z) &= \ln z + \psi \left( \frac{1}{2} + \frac{1}{z} \right), \quad B_{\phi, i, *} = \frac{\hbar}{4 D e} \tau_{\phi, i, *}^{-1} \tag{1}
    \label{WL_BLG}
    \end{align}
    
    where the characteristic fields $\quad B_{\phi, i, *}$ are expressed in terms of various characteristic scattering times. $B_{\phi}$ is related to the phase coherence time $\tau_{\phi}^{-1}$, $B_{i}$ is related to the intervalley scattering time $\tau_{i}^{-1}$, and $B_{*}$ to a combined scattering time $\tau_{*}^{-1}$, with 
    \begin{equation}
    \nonumber
    \tau_*^{-1} = \tau_i^{-1} + \tau_w^{-1},
    \end{equation}
    where $\tau_{w}$ is the intravalley warping time combined with the time of chirality breaking \cite{PhysRevLett.98.176805}. 

Magnetoconductance measurements were performed in a small field range ($\SI{-0.3}{T}$ to $\SI{0.3}{T}$) before and after CuPc deposition as shown in Figure \ref{fig:WL}. Measurements were carried out at two different temperatures, $\SI{1.5}{K}$ and $\SI{5}{K}$ to corroborate the decrease of the quantum interference effects with the increasing temperature. The data collected was then fitted to Equation \ref{WL_BLG} to extract the three scattering times ($\tau_{\phi}^{-1}, \tau_{i}^{-1}, \tau_{w}^{-1}$) at these temperatures, represented in Figure \ref{fig:WL}. We can see that $\tau_{\phi}$ decreases with temperature as expected.    
It is also noticeable that the addition of CuPc has an impact on all the characteristic times. It decreases the phase coherence time $\tau_{\phi}$, that quantifies the typical travel time of an electronic wave packet before loosing its phase coherence. From the bahavior of $\tau_{\phi}$, we can infer the coupling of the interfering electrons to the environment, composed in general by other electrons, phonons, electromagnetic fluctuations, magnetic impurities, etc. Oftentimes, $\tau_{\phi}$ is related to inelastic collisions, however in some cases, phase breaking processes do not involve an exchange of energy between the interfering electron and its environment, leading to a temperature independent value. This is the case for example of scattering with paramagnetic impurities that create a spin flipping of the interfering electron \cite{Akkermans1996-cf}. In our case, both the deposition of CuPc and temperature have an impact on $\tau_{\phi}$, showing that the phase breaking occurs in scattering events that go beyond energy-conservation processes such as spin-flip scattering with the paramagnetic CuPc molecules. The phase coherent time in the original BLG/h-BN device, $\tau_{\phi}=(2.3\pm 0.2)\times 10^{-11}$ s dropped by a factor of 2 after the deposition of the molecules, $\tau_{\phi}=(1.1\pm 0.1)\times 10^{-11}$ s, remaining 3 orders of magnitude larger than $\tau_e$. We deduce therefore that the motion of the electrons over a phase relaxation time is not ballistic. In this regime, the length of the electronic trajectories over the time period $\tau_{\phi}$ can be understood as the sum of short trajectories in random directions, given by $l_{\phi}=\sqrt{D\tau_{\phi}}$ \cite{Datta1995-it}, known as the phase coherent length, that in our sample changes from $460\pm30$ nm to $290\pm20$ nm after the deposition of the molecules. This places our device in a mesoscopic regime at all times, as the distance between the measured electrodes, 0.98 and 5.7 $\mu$m, is comparable to $l_{\phi}$. It is interesting to note that the electronic trajectories keep their phase coherence even when scattered across different CuPc grains, that have an average size of $42\pm0.7$ nm.

From our fit to the theory of weak localization for BLG, we also deduce the intervalley scattering time $\tau_i$, the time it takes for carriers to scatter from one valley to the other.  We find a value of the order of $10^{-10}$ s, about 1 order of magnitude larger than the values reported in the past for BLG on SiO$_2$ \cite{PhysRevLett.98.176805}. The larger values reported here can be consequence of the h-BN isolating the BLG from scatterers from the substrate and therefore disfavoring intervalley scattering. We find that $\tau_i$ decreases after the deposition of the molecules, from $\tau_i=(5.9\pm0.79)$ to $(2.0\pm0.33)\times 10^{-10}$ s. In general, intervalley scattering requires sharp defects, such at the edges of the sample, as it takes a large momentum transfer to change electrons from one valley to the other \cite{PhysRevLett.103.226801}. The observed reduction of $\tau_i$ after the CuPc deposition, indicates an increase in scattering between the K and K' valleys due to the presence of the molecules. We can estimate the intervalley diffusion length, L$_i=\sqrt{D\tau_i}$, that measures a characteristic distance between defects that can significantly change the momentum of the carriers \cite{PhysRevLett.98.176805}. This characteristic length changes from L$_i=2.3\pm0.2$  $\mu$m to $1.3\pm0.1$ $\mu$m after CuPc, indicating that carriers remain in the same valley through collisions that extend over multiple grains of CuPc. 

Finally, our fit to the magnetoresistance provides information about a very interesting elastic process that is strong in BLG compared to its monolayer counterpart, the warping on the energy spectrum \cite{PhysRevLett.98.176806}. The nonlinear least squares algorithm used for the fitting, allowed us to extract the value of the characteristic time associated with this scattering mechanism, brought by the third term in equation \ref{WL_BLG}. Trigonal warping refers to a three-fold perturbation of the circular iso-energetic lines that are characteristic of BLG (as well as monolayer graphene) near each K and K' valley. It occurs at high energies, as the band structure mimics the symmetry of the crystal lattice, where the consideration of nearest-neighbor hopping leads to a complex momentum dependence \cite{McCann_2013}. In BLG and bulk graphite, trigonal warping also appears at low energies as a result of the skew interlayer hopping between carbon atoms that do not lie directly above or below each other and are therefore not coupled through strong interlayer hoppping. Such oblique coupling is quantified by the tight-binding parameter $\gamma_3$, and has an important influence on the band structure at very low energies. At a threshold energy of $\approx$1 meV, the isoenergetic lines go through a Lifshitz transition, breaking into four pockets, corresponding to four Dirac cones, one at $\Gamma$ and three others at the corners of a triangle \cite{McCann_2013}. Above this energy, the effect of the $\gamma_3$ coupling can be treated as a perturbation, and creates a trigonal deformation of the circular single-connected Fermi line leading to an asymmetry of the energy dispersion: $\epsilon(\textbf{K},\textbf{p})\neq\epsilon(\textbf{K},-\textbf{p})$, with $\epsilon(\textbf{K},\textbf{p})=\epsilon(\textbf{K'},-\textbf{p})$ and \textbf{p} measured with respect to the center of the valley (see inset of Figure \ref{fig:WL}) \cite{PhysRevLett.98.176806}. Such deformation of the Fermi line has been experimentally observed through quasiparticle scattering experiments that show a triangular shape of the Fermi surface near K and K' at energies up to 112 meV \cite{PhysRevB.101.161103}. The \textbf{p}\textrightarrow-\textbf{p} asymmetry of the dispersion at each K and K' valley, creates a dephasing of electronic trajectories as those discussed at the introduction, and is characterized by the characteristic time $\tau_w$ \cite{Kechedzhi2007-tc}. Trigonal warping can therfore potentially destroy the manifestation of chirality in the localization properties of BLG, leading to the suppression of weak localization \cite{PhysRevLett.98.176806, Kechedzhi2007-tc}. 
When there is a strong scattering due to trigonal warping ($\tau_w\rightarrow 0$) and no intervalley scattering ($\tau_i\rightarrow\infty$), the third term in equation \ref{WL_BLG} disappears while the first two cancel each other, resulting in a zero magnetoconductance and a complete vanishing of weak localization. If instead there is some intervalley scattering, the second term in equation \ref{WL_BLG} is smaller than the first term, and there is some weak localization. This is the case in our experiment as well as in previous reports \cite{PhysRevLett.98.176805}. In our work, we find that the intervalley scattering time $\tau_i$ is 10 times larger than the phase coherent time $\tau_{\phi}$. Also, the intervalley diffusion length L$_i$ is about 2 times larger than the length of the measured junction (1 $\mu$m). We still find, however, a weak localization correction at zero fields (see Figure \ref{fig:WL}a)), indicating that there are some intervalley scattering events across the measured junction. Interestingly, the addition of the CuPc molecules enlarges the weak localization correction (see Figure \ref{fig:WL}a), reflected in a $\tau_w$ that is 2.5 times larger than the one measured in the original BLG/h-BN device, $(25\pm 2)\times10^{-14}$ vs $(9.7\pm 4)\times10^{-14}$ s. These values correspond to energies of 26-68 meV, which coincides with the energy scale where trigonal warping effects are important in BLG. The increase in $\tau_w$ suggests that the addition of the molecules has a tendency to bring back the manifestation of the carrier chirality in the localization of BLG, by reducing trigonal warping-related scattering events.



 \section{Conclusion}
 We have characterized a heterostructure of CuPc/BLG/h-BN, where the crystallinity of the CuPc thin film on the BLG/h-BN was identified through the analysis of a TEM-compatible control sample, studied through 4D-STEM, a diffraction space imaging technique that minimizes the radiation damage of sensitive samples such as CuPc molecules. This technique provided an estimate of the size of the crystalline grains of CuPc on our transport sample, of about $\approx 42$ nm. Electronic transport measurements at low temperatures before and after the deposition of the molecules revealed an important charge transfer of electrons from the BLG/h-BN to the CuPc molecules, of $3.6\times 10^{12}$/cm$^2$ corresponding to $\approx$ 67 electrons transferred to each grain of CuPc. Unexpectedly, we found a preservation of the mobility of the BLG/h-BN device upon the deposition of the molecules, degraded only by 13\%. The width of the Dirac point lightly increased, that we interpreted as the CuPc adding some few charged impurities to the device, about $\approx$ 2 impurities per grain of CuPc. Most importantly, our weak localization measurements disclosed that while the deposition of the molecules creates more scattering events that break the phase coherence of the carriers and that transfer electrons from one valley to the other, 
 surprisingly, it triggers an improvement of the trigonal warping time $\tau_{w}$. This corresponds to a reduction of the scattering associated to the trigonal deformation of BLG's Fermi surface near K and K'. As trigonal warping is known to destroy weak localization effects, our results suggest that the deposition of the CuPc molecules tends to restore the manifestation of the chirality of the carriers in the localization properties of BLG.

\section*{Supplementary Material}
Supplemental materials include the Raman scan of the CuPc/BLG/h-BN device, details on the analysis of the 4D-STEM data for the extraction of the size of the crystalline grains of CuPc, fits to the conductivity of the device that allow us to deduce the value of the Dirac point and its uncertainty, as well as the back gate dependence of the mobility and mean free path in the device before and after the deposition of CuPc.

\subsection*{Raman scan of the CuPC/BLG/h-BN device} 
\begin{figure}[h!]
\includegraphics[width=0.5\textwidth]{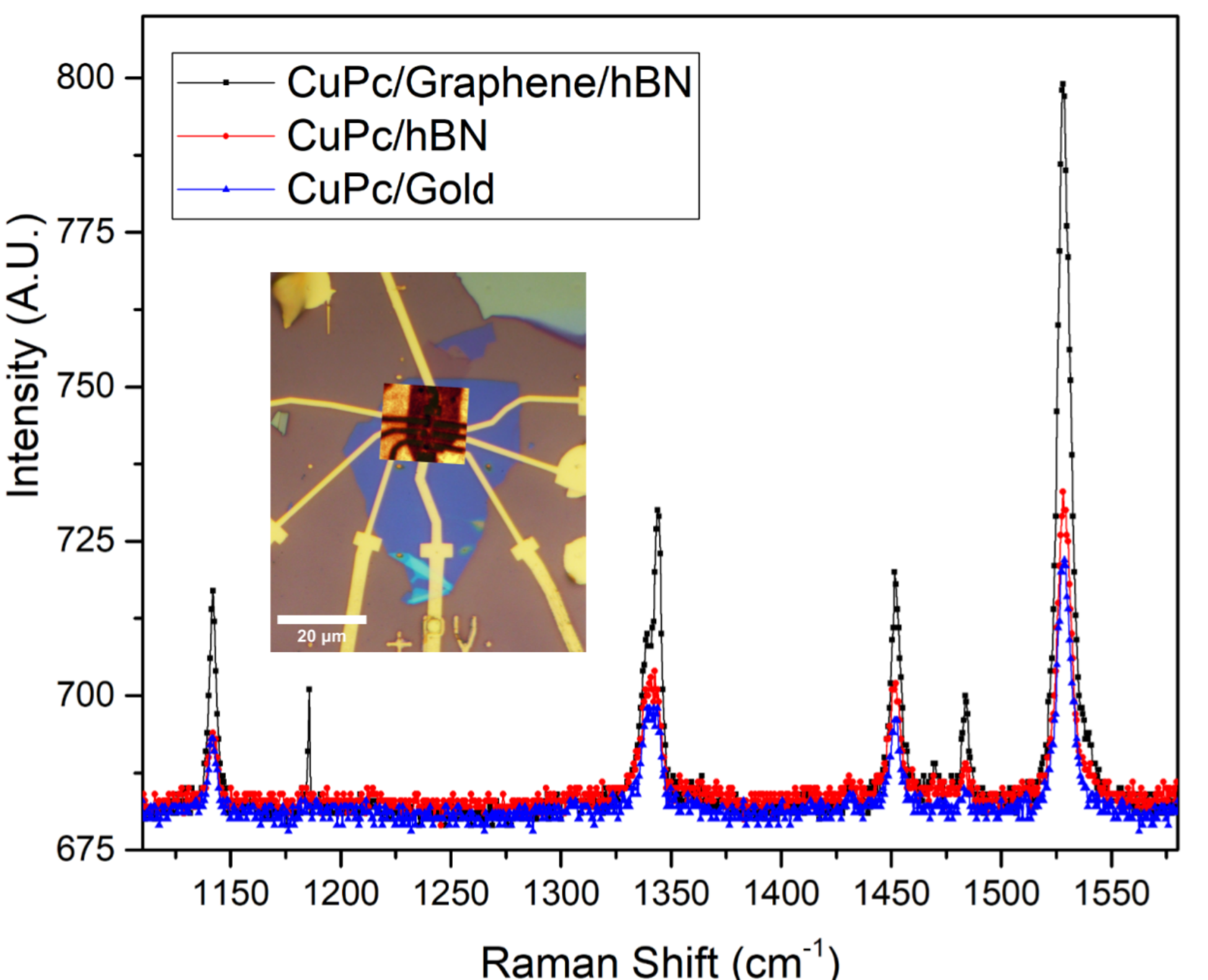}
\caption{Raman spectra of the CuPc/BLG/h-BN device. A change in intensity for all of the peaks depending on the substrate can be observed. Inset: virtual image constructed from the intensity of the 1530 cm\textsuperscript{-1} peak as the laser is scanned across an area of the device.} 
\label{fig:RamanCuPc}
\end{figure}

Figure \ref{fig:RamanCuPc} shows the Raman spectra of the CuPc/BLG/h-BN device. The characteristic peaks at 1185 and 1342 cm\textsuperscript{-1} are associated with the B\textsubscript{1g} vibrational mode. Additional peaks at 1142, 1452, and 1531 cm\textsuperscript{-1} correspond to the B\textsubscript{2g} mode, as previously reported \cite{ling_fang_lee_araujo_zhang_rodriguez-nieva_lin_zhang_kong_dresselhaus_2014}.
The Raman peaks were observed to be more pronounced on CuPc/hBN compared to the CuPc/BLG/hBN region on the device. Following CuPc deposition, Raman peaks associated with the underlying BLG crystal were no longer detectable.

\subsection*{Numerical analysis of the virtual bright-field image obtained from 4D-STEM}

\begin{figure*}
\includegraphics[width=\textwidth]{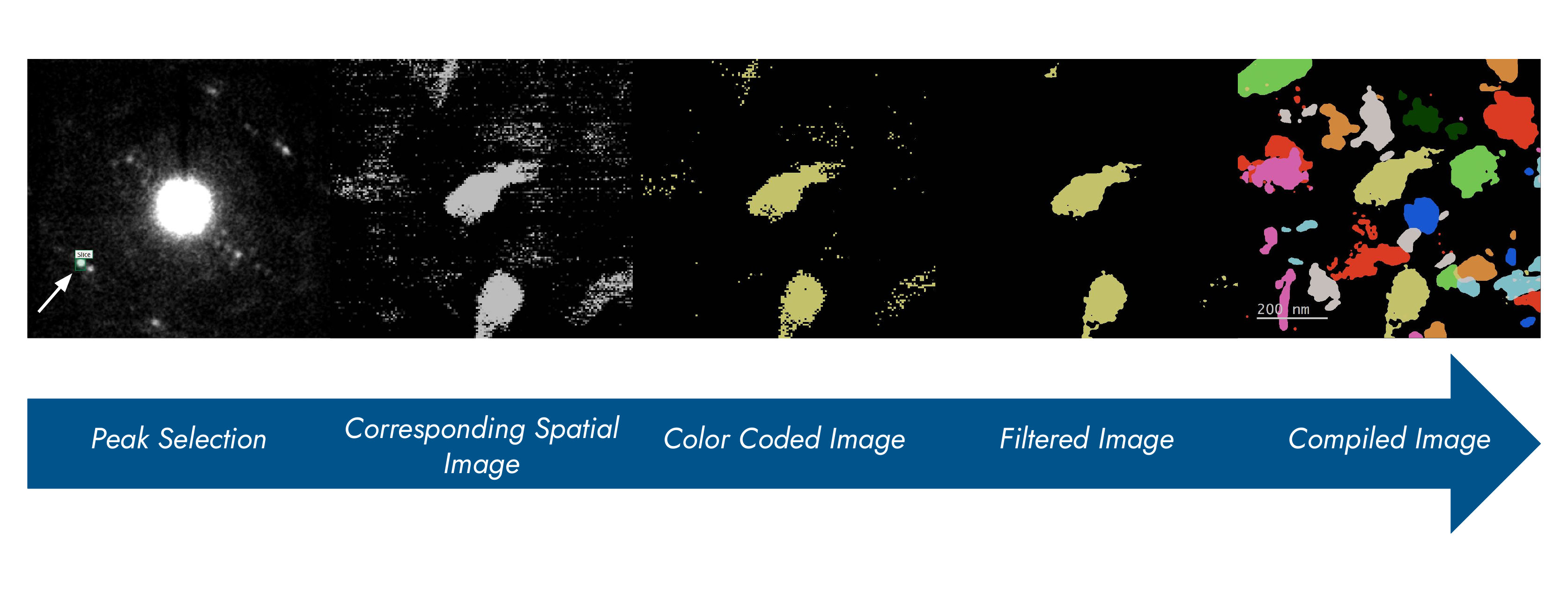}
\caption{Workflow for the 4D-STEM data analysis. The process starts with the peak selection of individual Bragg peaks via the manipulation of the virtual objective aperture (slice) as indicated by the white arrow. Each pixel in the corresponding spatial image is a function of its position in real space  and the intensity of the area covered by the slice in reciprocal space. The spatial images are then color coded and filtered with a Gaussian blur filter to remove the noise before forming compiled image.}
\label{fig:4DSTEM_Workflow}
\end{figure*}

\begin{figure}[h!]
\includegraphics[width=0.5\textwidth]{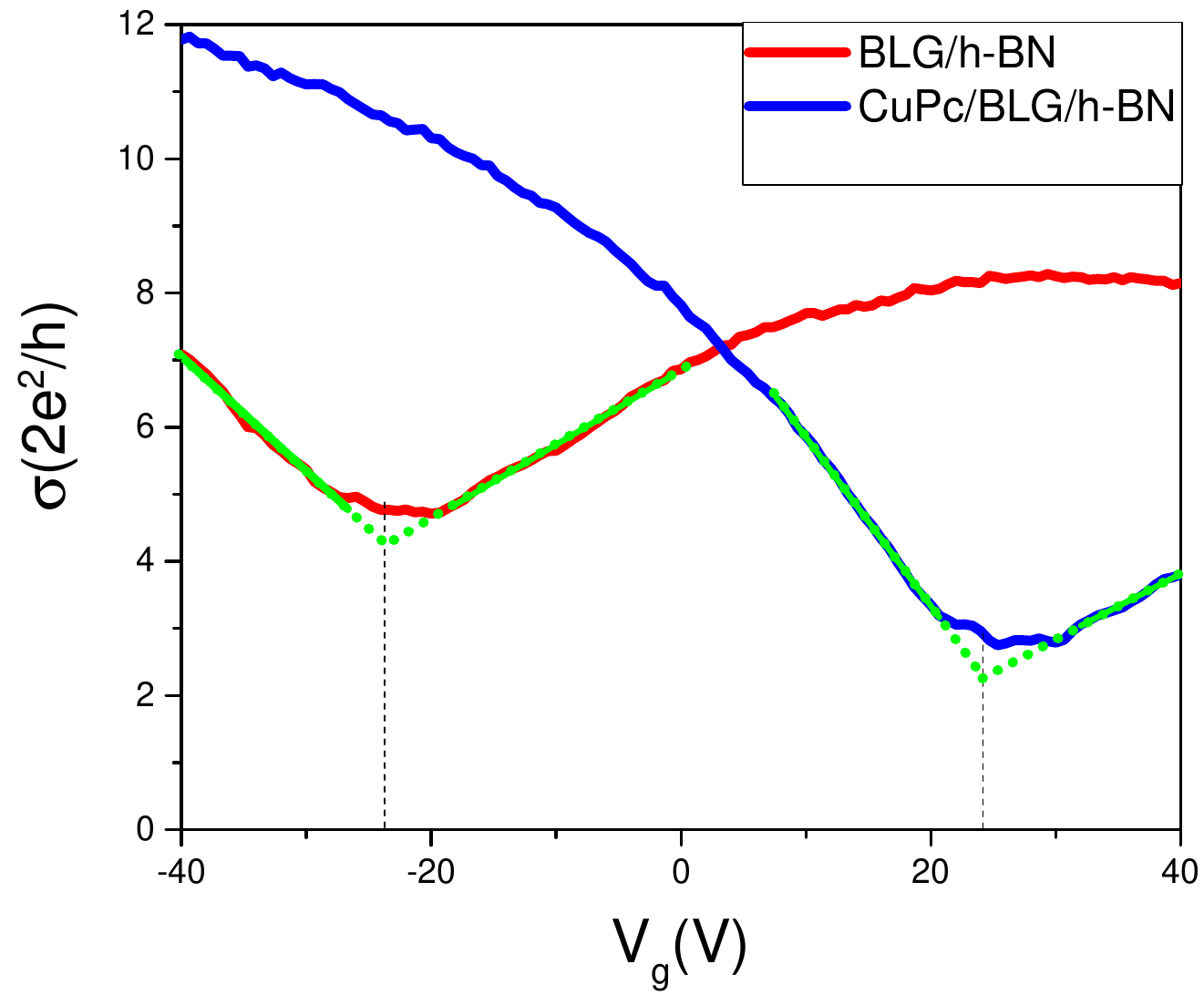}
\caption{Conductivity of the BLG/h-BN device before (red) and after (blue) the deposition of the CuPc molecules. Fits close to the Dirac point are shown in green, with guides to the eye (pointed lines) that show a Dirac point at -24 V before and 24 V after the deposition of CuPc.} 
\label{fig:sigma}
\end{figure}

\begin{figure}[h!]
\includegraphics[width=0.5\textwidth]{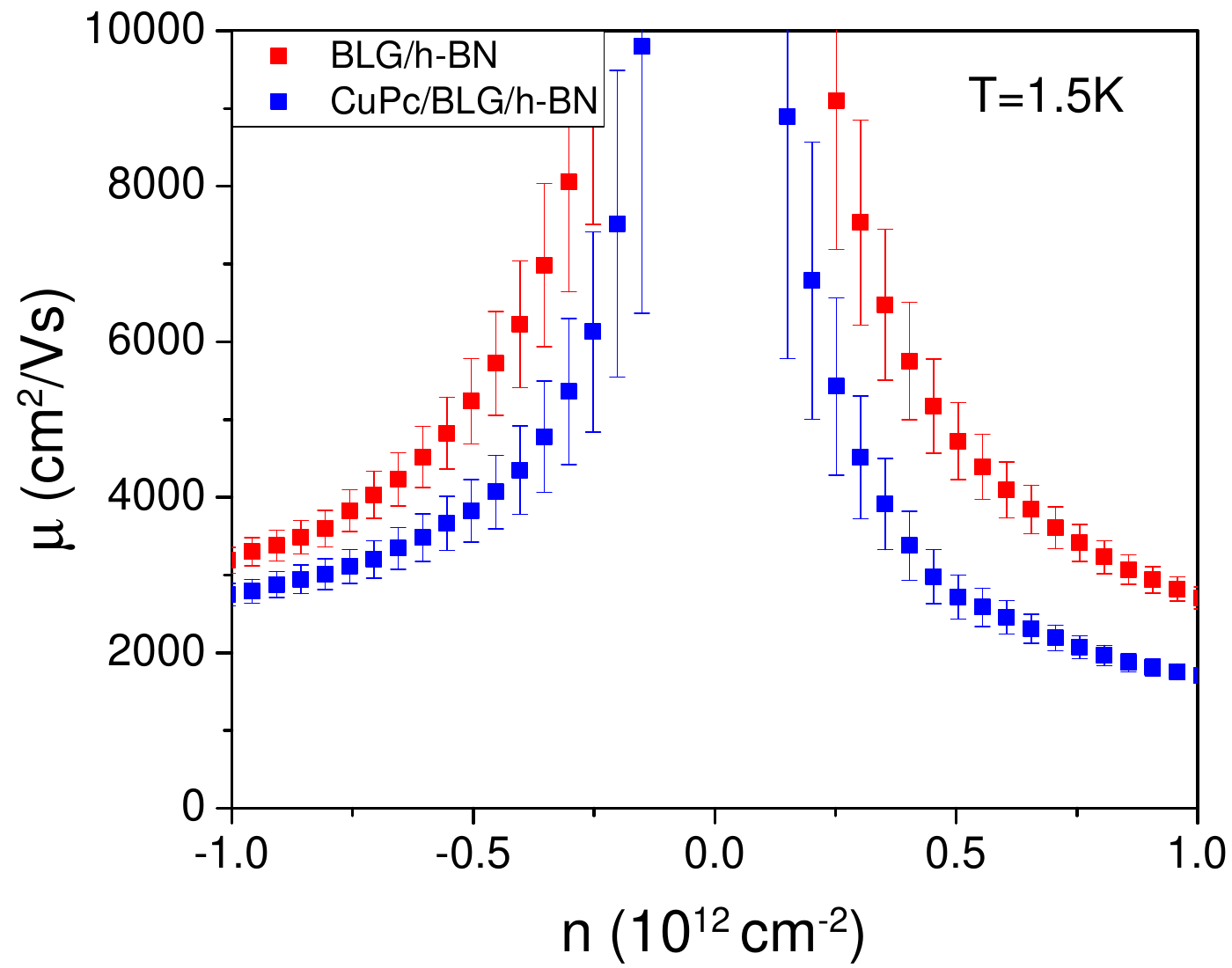}
\includegraphics[width=0.5\textwidth]{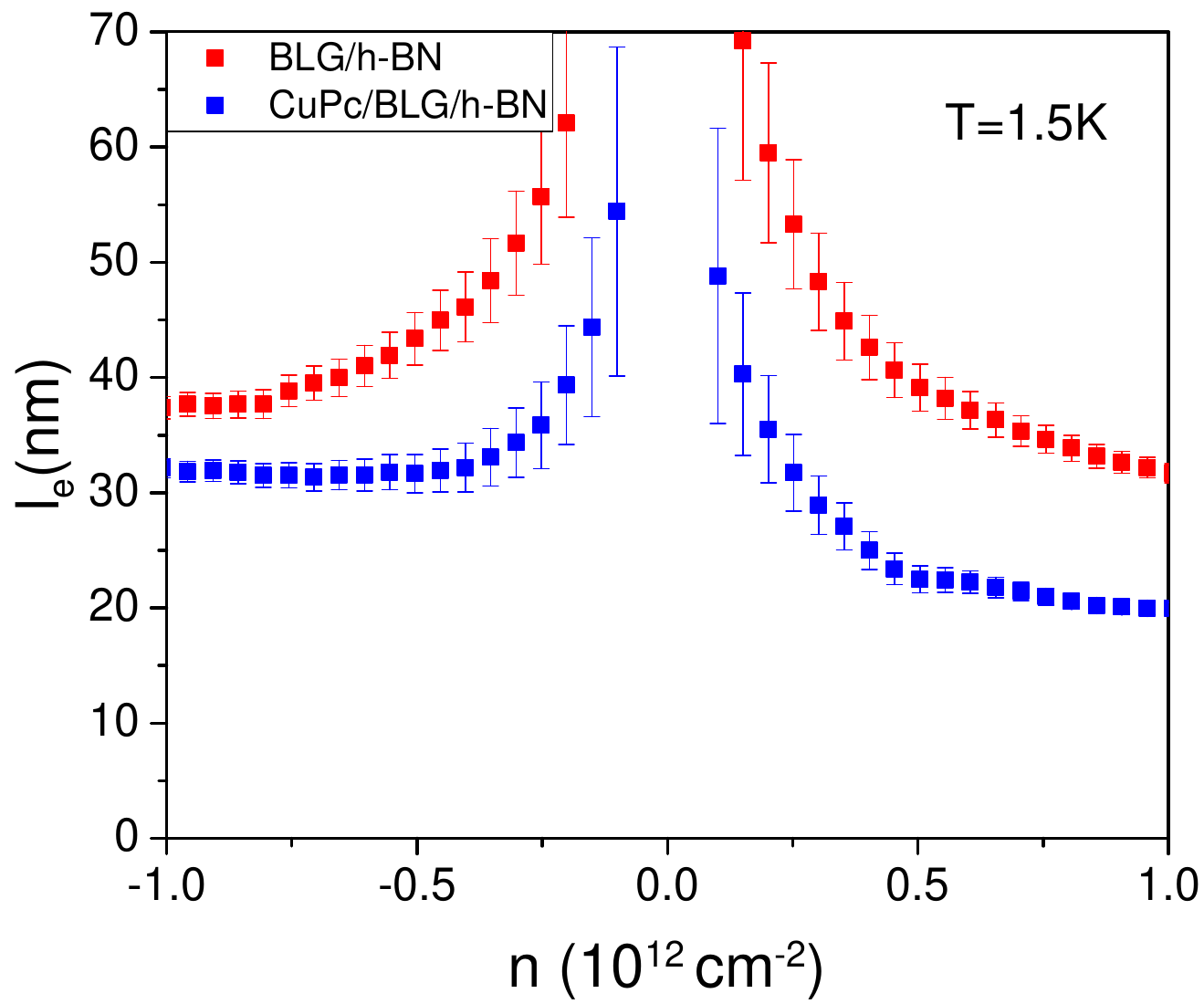}
\caption{Mobility (top) and mean free path of the BLG/h-BN device before and after the deposition of CuPc, measured at 1.5 K} 
\label{fig:Mu_mfp}
\end{figure}

4D-STEM data was collected on Thermo Fisher Spectra 300C scanning transmission electron microscope with an X-CFEG electron source operated at 300 kV. A 5 $\mu$m C2 aperture was used in microprobe mode to create a beam with a pseudo-parallel beam with a convergence angle of 0.1 mrad. Ceta-D (Thermo Fisher Scientific) was used with a dwell time of 10 ms. The electron beam was rastered across the sample for 22500 probe positions which translates to a 150 $\times$ 150 pixel resolution in real space. Simultaneously, a 256 $\times$ 256 pixel diffraction pattern was captured at each probe position by the detector.

Each of the virtual images of CuPc grains had a Gaussian blur filter applied to remove the noise followed by a color ramp to colorize the individual grains. These images were then converted to a two-dimensional array with each element representing the color of the pixel. The length of consecutive pixels with the same intensity was summed for each row of pixels and divided by the number of lengths to obtain the average width of a grain. Similarly, the code was processed on the transposed image to obtain the average height. The final grain size is an average these two dimensions. The 4D-STEM data process is illustrated in Figure \ref{fig:4DSTEM_Workflow}.

\subsection*{Determining the Dirac point from the conductivity}
Figure \ref{fig:sigma} shows the conductivity before and after the deposition of the CuPc molecules in units of e$^2$/h. We can see by naked eye, from the slope of the conductivity away from the Dirac point, a preserved mobility after the deposition of the molecules. The conductivity was considered to be linear with the gate voltage (or electronic density), following the model that considers charged impurities to be the main scattering mechanism in graphene \cite{Adam2007-fn, Chen2008-dt}. This allowed us to determine the gate voltage expected at the charge neutrality point, by drawing guides for the eye that elongated the fits, as done in \cite{Chen2008-dt}. The errors of these fits allowed has to determine an uncertainity for the Dirac point V$_D$, $\Delta V_D=\pm0.7$ V. We can also observe an increase in the width of the Dirac point, from 8.6$\pm$0.7 V in the BLG/h-BN device to 10$\pm$0.7 V after the deposition of CuPc as explained in the main text. 

The uncertainty in V$_D$ was propagated to find the uncertainities on the charge transfer, mobility, Fermi energy, mean free path, Fermi wave vector, elastic scattering time, phase coherent length and intervalley diffusion length, as reported in the main text.

\subsection*{Mobility and Mean free path before and after the deposition of CuPc}
Figure \ref{fig:Mu_mfp} shows the calculated mobility and mean free path of the BLG/h-BN device before and after the deposition of the CuPc. We observe an asymmetry in the mean free path and mobility for the electrons and holes. This can be attributed to the intrinsic electronic properties of the bilayer graphene, characterized by an electron-hole asymmetry that results from one of the interlayer hopping parameters, measured in experimentally in the past \cite{PhysRevB.84.085408}. The assymetry can also be result of a spatially inhomogeneous potential distribution of charged impurities \cite{Adam2007-fn}. The error bars correspond to the error propagation of $\Delta V_D$.
\begin{acknowledgments}
The primary funding for this work was provided by the U.S. Department of Energy, Office of Science, Office of Basic Energy Sciences under contract DE-SC0018154. The deposition of the molecular thin film was funded by the Cal. State. Long Beach and the Ohio State University Partnership for Education and Research in Topological Materials, a National Science Foundation PREM, under Grant No. 2425133. 4D-STEM measurements were possible thanks to BioPACIFIC, an NSF Materials Innovation Platform (DMR-1933487). Preliminary TEM measurements by Y.C., H.C and B.C.R. were funded by the National Science Foundation, DMR-1548924 (STROBE). K.W. and T.T. acknowledge support from the JSPS KAKENHI (Grant Numbers 21H05233 and 23H02052) , the CREST (JPMJCR24A5), JST and World Premier International Research Center Initiative (WPI), MEXT, Japan for the h-BN growth. We would like to acknowledge very insightful discussions with Francisco Guinea, Maria Jose Calderon, Andreas Bill, Sophie Gu\'eron and H\'el\`ene Bouchiat as well as invaluable advice from David Warren from Oxford Instruments. D.D,  M.M and E.C.-O. would like to acknowledge the Google American Physical Society Bridge Program, M.K.D.D.M. the Google summer assistantship at CSULB. A.M. would like to acknowledge the Keung Luke, Charles Roberts and Richard Whiteley Endowed Scholarship at CSULB.
\end{acknowledgments}
\bibliography{Ojeda-AristizabalBibCuPc-BLG}

\end{document}